\documentclass{article}[12pt]
\usepackage{graphicx}
\usepackage{hyperref}
\usepackage{amssymb}
\usepackage{multirow}
\usepackage{caption}
\usepackage{subcaption}
\usepackage[left=1.5cm,right=1.5cm,top=1.5cm,bottom=1.5cm]{geometry}
\usepackage{authblk}
\usepackage{amsmath}
\begin{document}
	\title{Exploring gravastar-like structures with strongly interacting quark matter shell in the framework of $f(Q)$ gravity under conformal symmetry}
	\author[1]{Debadri Bhattacharjee\thanks{debadriwork@gmail.com}}
	\author[2]{Pradip Kumar Chattopadhyay\thanks{pkc$_{-}76$@rediffmail.com}}
	\affil[1,2]{IUCAA Centre for Astronomy Research and Development (ICARD), Department of Physics, Cooch Behar Panchanan Barma University, Vivekananda Street, District: Cooch Behar, \\ Pin: 736101, West Bengal, India}
	\maketitle

\begin{abstract}
In this work, we investigate gravastar-like structures in static and spherically symmetric space-time within the framework of $f(Q)$ gravity coupled with conformal symmetry. We have modified the conventional gravastar model by introducing a strongly interacting quark matter shell which maintains the apex of causal limit through the EoS, $p=\rho-2B_{g}$, where, $B_{g}$ is the bag constant. Non-singular and non-vanishing solutions for the interior and shell regions are obtained, respectively. We have used the Israel junction condition to evaluate the mass of the thin shell for different choices of characteristic radii. Interestingly, the mass of the shell is independent of the matter distribution in the shell region. We found that for radii 9.009, 10.009 and 11.009, the mass increases as $1.80,~1.95$ and $2.28~M_{\odot}$. The physical features, such as, proper length, energy and entropy of the shell region are studied within the parameter space. Surface redshift calculations were used to validate the proposed model.     

\end{abstract}


\section{Introduction}\label{sec1} The nomenclature Gravitational Vacuum Condensate Stars, abbreviated as Gravastars, was first coined by P. O. Mazur and E. Mottola \cite{Mazur,Mazur1,Mazur2} in the beginning of twenty first century. Till then, from the perspective of classical gravity, Black Hole (BH) carried an element of unresolvable mystery in terms of the central singularity. As a demarcation of the end stage of stellar evolution, BHs are not only interesting because they are the direct consequence \cite{Schwarzschild} of Einstein's General theory of Relativity (GR), but also due to their immense entropy which originates from quantum mechanics \cite{Hawking}. However, the question of singularity persists. Roger Penrose demonstrated that the formation of BH singularities can occur even from collapsing matter that lacks perfect spherical symmetry \cite{Penrose}. Subsequently, observational evidence confirming the existence of these massive, compact objects, including BHs, was provided by the work of Andrea Ghez \cite{Ghez} and Reinhard Gillessen \cite{Gillessen}. Notably, Penrose, Ghez, and Gillessen were jointly awarded the 2020 Nobel Prize in Physics for their groundbreaking contributions to our understanding of BHs. However, there are several problems in accepting BHs as the ultimate fate of gravitational collapse. Following the works of Schwarzschild \cite{Schwarzschild} and Penrose \cite{Penrose}, it is evident that there is an inescapable central singularity where GR breaks down. Further, the Hawking radiation associated with a BH shows that the Hawking temperature $(T_{BH}=\frac{\hbar c^{3}}{8\pi G k_{B}M})$ is inversely proportional to mass \cite{Hawking1}. This situation is paradoxical in its own as the inverse dependence of $T_{BH}$ on $M$ gives rise to negative heat capacity which in turn destabilises a system. Hence, an alternative approach was essential and gravastar model \cite{Mazur,Mazur1,Mazur2} offered the solution. The gravastar model initially consisted of five distinct layers, but Visser and Wiltshire \cite{Visser} later simplified it to a widely accepted three-layer structure comprising an interior region, a thin shell, and an exterior region. Each region is described by a specific equation of state (EoS). The interior region is characterised by a de-Sitter condensate phase with negative pressure or positive energy density ($p=-\rho$), corresponding to a positive cosmological constant introduced by Einstein to support a static universe based on Mach's principle. The negative pressure implies that the interior generates an outward repulsive pressure acting on the thin shell. The shell, composed of an ultra-relativistic stiff fluid, follows the EoS $p=\rho$, consistent with the Zel'dovich criterion \cite{Zeldovich,Zeldovich1} for a cold baryonic universe. In this region, the speed of sound matches the speed of light, ensuring maximum causal consistency. The shell is thin but has a finite thickness. The exterior region is a vacuum flat spacetime, described by the EoS $p=\rho=0$. 

Mazur and Mottola in their work \cite{Mazur,Mazur1,Mazur2} posited gravastars as a alternative explanation to BHs. Soon after their monumental hypothesis, Chirenti and Rezzolla \cite{Chirenti} examined two key questions regarding gravastars: whether they can form stable configurations and how to distinguish them from BHs. Carter \cite{Carter} developed a gravastar model with a generalised Schwarzschild–(anti)-de Sitter or Reissner–N\"ordstrom exterior. Ray et al. \cite{Ray} demonstrated that for a gravastar to be physically viable within GR, the presence of pressure anisotropy and adherence to specific energy conditions are essential. Horvat and Ilijic \cite{Horvat} formulated the energy conditions for gravastars and derived a strict limit on surface compactness necessary to satisfy the dominant energy condition at the thin shell. DeBenedictis et al. \cite{DeBenedictis} constructed a stable gravastar model characterised by continuous pressure and density. Further, Cattoen {\it et al.} \cite{Cattoen} analysed gravastar models with pressure anisotropy without relying on the thin shell approximation. Models of gravastars have also been explored in higher dimensions. Rahaman et al. \cite{Rahaman} proposed a gravastar configuration in $(2+1)$ dimensions, while in another study, Rahaman et al. \cite{Rahaman1} extended the analysis to D-dimensional Einstein gravity. Ghosh et al. \cite{Ghosh} developed a gravastar model in $(3+1)$ dimensions using the Karmakar class-I condition. Bhar \cite{Bhar} studied higher-dimensional charged gravastars in the context of conformal motion. Bilic et al. \cite{Bilic} explored an alternative gravastar model by replacing the de-Sitter interior with a Born-Infeld phantom field, linking the model to low-energy string theory. Lobo \cite{Lobo}, inspired by cosmic expansion, examined a gravastar solution governed by a dark energy EoS, analysing several dark energy stellar parameters. Gravastars have also been studied under the light of cylindrical symmetry. For example, Bhattacharjee et al. have employed cylindrical space-time to study a generalised gravastar structure and compute the mass limit of thin shell \cite{Bhattacharjee}. Furthermore, Bhattacharjee and Chattopadhyay \cite{Bhattacharjee2} investigated charged gravastar models within the framework of general cylindrically symmetric spacetime.

The modified theories of gravity were introduced to counter the shortcomings of GR. While Einstein's GR remains the most well-established theory for explaining gravitational phenomena, observational evidence of cosmic acceleration, dark energy, and dark matter presents significant challenges to GR's theoretical predictions \cite{Riess}-\cite{Tegmark}. The necessity of extending GR through modified gravity theories has been extensively discussed in \cite{Nojiri}. Generalising the Einstein-Hilbert action has led to the development of several alternative gravitational models, including $f(R)$ gravity (where $R$ is the Ricci scalar) \cite{Capozziello,Elizalde}, $f(\mathcal{T})$ gravity (where $\mathcal{T}$ is the torsion scalar) \cite{Bamba}, $f(R,\Box R,T)$ gravity (where $\Box$ is the d'Alembert operator and $T$ is the trace of the energy-momentum tensor) \cite{Houndjo}-\cite{Ilyas}, $f(R,T)$ gravity \cite{Harko} and so on. Moreover, shortly after the development of GR, Weyl \cite{Weyl} introduced the first modifications by incorporating higher-order invariants into the Einstein-Hilbert action, aiming to unify gravitation with electromagnetism. Subsequent modifications to GR have been categorised into two primary geometric extensions: (i) Curvature-based extensions that include theories such as $f(R)$, $f(T)$, and $f(R,T)$ gravity, and (ii) Torsion and non-metricity based extensions that involve theories such as $f(Q)$ gravity, which generalise GR through torsion and non-metricity \cite{Einstein}. Theories based on torsion and non-metricity, which are dynamically equivalent to GR, are known as the Teleparallel Equivalent of General Relativity (TEGR) \cite{Hayashi,Sauer} and the Symmetric Teleparallel Equivalent of General Relativity (STEGR) \cite{Nester,Jimenez,Jimenez1}. TEGR is constructed in a flat spacetime with torsion, where the tetrads and spin connections serve as the fundamental variables. The requirement of vanishing curvature and non-metricity tensors constrains the spin connection, allowing the use of the Weitzenb\"ock connection. Under this connection, the spin connection terms are eliminated, leaving the tetrads as the core variables. This configuration is treated as a gauge choice within TEGR. Importantly, this choice does not affect the physical outcomes of the theory, as any other selection consistent with the teleparallelism condition would lead to the same action, up to a boundary term.

In STEGR, gravity is defined by non-metricity rather than curvature and torsion. A specific gauge, known as the coincidence gauge, can be adopted, reducing the metric tensor to the sole fundamental variable. An extension of STEGR is the $f(Q)$ gravity theory, which parallels the $f(R)$ gravity framework \cite{Jimenez,Heisenberg}. Hohmann et al. \cite{Hohmann} examined gravitational wave polarisation and propagation velocity in Minkowski space within the $f(Q)$ gravity framework. Soudi et al. \cite{Soudi} further demonstrated that gravitational wave polarisation significantly influences the strong-field properties of modified gravity theories. The $f(Q)$ framework has been applied to various phenomena, including late-time cosmic acceleration \cite{Lazkoz}, black hole solutions \cite{Ambrosio,Heisenberg1,Calza}, bouncing cosmologies \cite{Bajardi}, the growth index evolution in matter perturbations \cite{Khyllep}, and observational constraints through $f(Q)$ parametrization \cite{Ayuso}. A comprehensive review of $f(Q)$ gravity is available in Ref. \cite{Heisenberg2}, with additional insights provided in Refs. \cite{Barros,Jimenez2,Anagnostopoulos,Flathmann,Heisenberg3}. 

A comprehensive and thorough study of any compact object requires looking beyond the conventional beliefs. In this situation, studying compact stellar structures in deviations from GR, i.e., in the framework of modified theories of gravity has become a very important outlet. Notably, gravastar models have also been extensively explored within the framework of various modified gravity theories. Das et al. \cite{Das} examined gravastar formation in $f(R,T)$ gravity, while Ghosh et al. \cite{Ghosh1} analysed the fundamental properties of gravastars within $f(T,T)$ gravity. Das et al. \cite{Das1} investigated gravastar structures with an exterior Schwarzschild–de Sitter solution in the context of $f(T)$ gravity. Sengupta et al. \cite{Sengupta} constructed gravastar models within the framework of braneworld gravity. Banerjee et al. \cite{Banerjee} adopted Finslerian geometry to develop gravastar configurations. Ghosh et al. \cite{BCP} explored gravastar structures in Rastall gravity, focusing on the influence of the Rastall parameter on the gravastar’s key characteristics. Yusaf \cite{Yusaf} examined the feasibility of charged gravastars within cylindrically symmetric spacetime in the framework of $f(R,T)$ gravity. Bhatti et al. \cite{Bhatti} investigated cylindrical gravastar-like structures under $f(G)$ gravity, while Pradhan et al. \cite{Pradhan} studied gravastar models within symmetric teleparallel gravity. Additionally, the gravastar framework has been explored under $f(G,T^{2})$ gravity in Ref. \cite{Bhatti1}. Recently, Bhattacharjee and Chattopadhyay \cite{Bhattacharjee3} have studied the existence of charged gravastars in Rastall gravity. Further, Mohanty et al. \cite{Sahoo} have employed the conformal symmetry to model charged gravastars in $f(Q)$ gravity. Various other studies have addressed different aspects of gravastar models \cite{Shamir, Rahaman2, Chan, Rocha, Lobo1, Horvat1, Turimov}. 

One of the most astonishing feature of a gravastar is the formation of shell region. The ultra-relativistic matter distribution in the shell sustains the negative pressure and maintains structural stability. In the pursuit of finding a new type of ultra-relativistic stiff fluid, we have come across the Strongly Interacting Quark Matter (SIQM). Recent work by Holdom et al. \cite{Holdom} introduced the idea that the bulk energy per baryon of quark matter composed solely of $u$ and $d$ quarks (udQM) is lower than that of SQM when flavour-dependent feedback on the QCD vacuum is considered. Consequently, udQM could be more stable than SQM, and under certain conditions, it may represent the ground state of baryonic matter at sufficiently high baryon densities. Zhang \cite{Zhang1} explored the implications of udQM in gravitational wave observations, proposing that the binary merger event GW170817 might have involved a $u$ and $d$ quark star (udQS). This hypothesis has inspired several phenomenological and experimental models \cite{Zhang1,Wang,Zhao,Xia,Acharya,Piotrowski}. Moreover, a realistic EoS for compact stars must account for the strong interactions among quarks and gluons, which are central to QCD. In this framework, Interacting Quark Matter (IQM) has become a leading model for describing the internal structure of SQS, accounting for the effects of strong interactions and the extreme densities present in compact stellar cores. The IQM model incorporates effects from perturbative QCD (pQCD) and colour superconductivity. Gluon-mediated corrections in IQM arise from pQCD interactions \cite{Farhi,Fraga,Fraga1}. Additionally, colour superconductivity emerges from the condensation of spin-0 Cooper pairs in the flavour space, forming two-flavour colour superconductivity (2SC) when $u$ and $d$ quarks pair, or colour-flavour-locked (CFL) phases when $u$, $d$, and $s$ quarks pair symmetrically \cite{Alford,Rajagopal,Lugones}. The 2SC+s phase includes $s$ quarks, extending the pairing structure. The unexpectedly high mass of the lighter companion in the GW190814 event \cite{Abbott} challenges non-interacting SQM and udQM models with existing bag constant values. This discrepancy suggests that strongly interacting quark stars (SIQSs), composed of IQM, might satisfy these mass constraints and better explain the properties of the lighter companion. The concept of IQM has driven renewed interest in modeling compact stars under extreme conditions \cite{Pretel,Errehymy1,Tangphati}, offering a more robust framework for understanding the internal structure of quark stars and their observational signatures. Further, following the work of Zhang and Mann \cite{Zhang}, we have noted that in the strong interaction regime, the SIQM describes the highest value of causality $(\frac{dp}{d\rho}=1)$, thereby resembling a stiff fluid within the gravastar shell. Consequently, we have modeled gravastar-like structures with an SIQM shell to explore the overall properties of gravastars, while maintaining the ultra-relativistic stiff fluid condition in the shell's EoS. 

The rest of the paper is organise as follows: in Section~\ref{sec2}, we have discussed the SIQM in details and in the strong interaction regime, we have obtained the SIQM EoS which is utilised in the shell region. Next, in Section~\ref{sec3}, we have presented the Einstein field equations in the framework of $f(Q)$ gravity considering static and spherically symmetric space-time coupled with conformal symmetry. Section~\ref{sec4} addresses the gravastar like structure comprising of SIQM shell. Here, we have solved the field equations in each of the three layers and obtained physically viable solutions within the parameter space. Section~\ref{sec5} deals with the junction condition associated with smooth matching of the interior and exterior solutions at the hypersurface $(r=R)$. In this section, we have also obtained the analytical expression for the mass of the thin shell. The numerical determination of the constant parameters are discussed in Section~\ref{sec6}. In Section~\ref{sec7}, we compute the mass of the thin shell for suitable combination of the model parameters. The key attributes of the model, such as proper length, energy and entropy analysis are evaluated in Section~\ref{sec8} and the stability of the present model is assessed via surface redshift, in Section~\ref{sec9}. Finally, we recap the principal conclusions in Section~\ref{sec10}. 
\section{Strongly interacting quark matter EoS}\label{sec2} 
We begin with the free energy $(\Omega)$ representing the super conducting quark matter within the framework of pQCD \cite{Alford1,Alford2,Weissenborn} in the following form \cite{Zhang}:
\begin{equation}
	\Omega=-\frac{\xi_{4}}{4\pi^{2}}\mu^{4}+\frac{\xi_{4}(1-a_{4})}{4\pi^{2}}\mu^{4}-\frac{\xi_{2a}\Delta^{2}-\xi_{2b}m_{s}^{2}}{\pi^{2}}\mu^{2}-\frac{\mu_{e}^{4}}{12\pi^{2}}+B_{g}, \label{eq1}
\end{equation}
In this context, $\mu_{e}$ and $\mu$ denote the average chemical potentials associated with electrons and quarks, respectively. In Eq.~(\ref{eq1}), the first term corresponds to the contribution from an unpaired free quark gas, while the second term arises from perturbative quantum chromodynamics (pQCD), specifically capturing the effects of one-gluon exchange interactions up to $\mathcal{O}(\alpha_{s}^{2})$. From a phenomenological standpoint, variations in the parameter $a_{4}$ from unity to significantly lower values incorporate higher-order corrections \cite{Alford2,Weissenborn}, where $a_{4}=1$, implies the absence of pQCD corrections. The third term accounts for the finite mass of the strange quark through $m_{s}$, and $\Delta$ represents the influence of colour superconductivity. The final term, characterised by the bag constant $B_{g}$, quantifies the effective energy difference between perturbative and non-perturbative QCD vacuum states, and its magnitude may depend on the choice of quark flavours \cite{Holdom}. Based on the analysis by Zhang and Mann \cite{Zhang}, it is observed that depending on the numerical values of the constant coefficients appearing in Eq.~(\ref{eq1}), distinct phases of quark matter emerge within the system, namely:
\begin{equation}
	(\xi_{4},\xi_{2a},\xi_{2b})=
	\begin{cases}
		\Bigg(\Big[(\frac{1}{3})^{\frac{4}{3}}+(\frac{2}{3})^{\frac{4}{3}}\Big]^{-3},1,0\Bigg)~~~~~\text{2SC Phase}\\
		(3,1,\frac{3}{4})~~~~~~~~~~~~~~~~~~~~~~~~~~~\text{2SC+s Phase}\\
		(3,3,\frac{3}{4})~~~~~~~~~~~~~~~~~~~~~~~~~~~\text{CFL Phase}
	\end{cases}
	\label{eq2}
\end{equation}
The physical variables associated with any system, {\it viz.}, pressure $(p)$, energy density $(\rho)$ and number densities related to quarks $(n_{q})$ and electrons $(n_{e})$ are expressed through the free energy $(\Omega)$ is the following form: 
\begin{equation}
	p=-\Omega,~~~~~~~~\rho=\Omega+n_{q}\mu+n_{e}\mu_{e}~~~~~~~~~~~n_{q}=-\frac{\partial\Omega}{\partial\mu},~~~~~~~~~~~n_{e}=-\frac{\partial\Omega}{\partial\mu_{e}}. \label{eq3}
\end{equation} 
To reduce the number of variables, compactify the parameter space and identify the magnitude of strong interaction \cite{Zhang}, we introduce a new parameter $\chi$ as,
\begin{equation}
	\chi=\frac{\xi_{2a}\Delta^{2}-\xi_{2b}m_{s}^{2}}{\sqrt{\xi_{4}a_{4}}}. \label{eq4}
\end{equation}
Utilising Eq.~(\ref{eq3}), we get
\begin{equation}
	n_{q}=\frac{\xi_{4}a_{4}}{\pi^{2}}\mu^{3}+\frac{\chi\sqrt{\xi_{4}a_{4}}}{\pi^{2}}2\mu, ~~~~~~~~~~~~~~~~n_{e}=\frac{\mu_{e}^{3}}{3\pi^{2}}, \label{eq5}
\end{equation}
and, 
\begin{equation}
	\rho=\frac{3\xi_{4}a_{4}}{4\pi^{2}}\mu^{4}+\frac{\mu_{e}^{4}}{4\pi^{2}}+\frac{\chi\sqrt{\xi_{4}a_{4}}}{\pi^{2}}\mu^{2}+B_{g}. \label{eq6}
\end{equation}
Using Eqs.~(\ref{eq1}), (\ref{eq3}) and (\ref{eq6}), we obtain a general form of EoS as,
\begin{equation}
	p=\frac{1}{3}(\rho-4B_{g})+\frac{4\chi^{2}}{9\pi^{2}}\Bigg(-1+\sqrt{1+3\pi^{2}\frac{(\rho-B_{g})}{\chi^{2}}}\Bigg). \label{eq7}
\end{equation}
Notably, Eq.~(\ref{eq7}) provides a unified framework to describe $2SC$, $2SC+s$ and $CFL$ phases for different values of $\xi$ as given in Eq.~(\ref{eq2}). Hence, Eq.~(\ref{eq7}) is termed the unified IQM EoS. There are two interesting observations here: (i) for $\chi\rightarrow0$, Eq.~(\ref{eq7}) retains the conventional form of the non-interacting quark matter EoS, and (ii) for $\chi\rightarrow\infty$, we get, 
\begin{equation}
	p=\rho-2B_{g}, \label{eq7a}
\end{equation} 
which describes the strongly interacting quark matter (SIQM) EoS. Further, we must elucidate that the inclusion of strong interaction effectively reduces the surface energy density from $4B_{g}$ to $2B_{g}$ and increases the sound velocity $(\frac{dp}{d\rho})$ associated with quark matter from $\frac{1}{3}$ to $1$. As a result, we move towards the regime of ultra-relativistic stiff fluid \cite{Zeldovich,Zeldovich1} and the apex of causality. 

\section{Einstein field equations in the context of $f(Q)$ gravity and conformal symmetry}\label{sec3}
In the context of symmetric teleparallel $f(Q)$ gravity, the action integral is formulated as follows:
\begin{equation}
	\mathfrak{S}=\int \sqrt{-g}d^{4}x\Bigg[\frac{1}{2}f(Q)+\lambda^{kij}_{l}R^{l}_{kij}+\tau^{ij}_{k}T^{k}_{ij}+\mathfrak{L_{m}}\Bigg], \label{eq8}
\end{equation}
where, $g=|g_{ij}|$ denotes the determinant of the fundamental metric tensor. The function $f(Q)$ depends on the non-metricity scalar $Q$. The terms $\lambda^{kij}_{l}$ and $\tau^{ij}_{k}$ act as Lagrange multipliers, 
$R^{l}_{kij}$ corresponds to the Riemann curvature tensor, and $T^{k}_{ij}$ represents the torsion tensor. Additionally, $\mathfrak{L_{m}}$ characterises the Lagrangian density associated with matter.

The non-metricity tensor is given in terms of the affine connection $(\Gamma^{k}_{ij})$ as
\begin{equation} 
 Q_{kij}=\nabla_{k}g_{ij}=\partial_{k}g_{ij}-\Gamma^{l}_{ij}g_{ij}-\Gamma^{l}_{ik}g_{jl}. \label{eq9}
\end{equation}
Here, $\nabla_{k}$ represents the covariant derivative. Furthermore, the affine connection can be decomposed into three distinct components:
\begin{equation} 
	\Gamma^{k}_{ij}=\epsilon^{k}_{ij}+K^{k}_{ij}+L^{k}_{ij}, \label{eq10}
\end{equation}
where, $\epsilon^{k}_{ij}$ is the Levi-Civita connection, derived from the metric tensor $g_{ij}$, and is given by:
\begin{equation} \epsilon^{k}_{ij}=\frac{1}{2}g^{kl}\Big(\partial_{i}g_{lj}+\partial_{j}g_{il}-\partial_{l}g_{ij}\Big). \label{eq11} 
\end{equation}
$K^{k}_{ij}$ represents the contorsion tensor and is defined as
\begin{equation} 
	K^{k}_{ij}=\frac{1}{2}T^{k}_{ij}+T_{(_{i}~k~_{j})}. \label{eq12} 
\end{equation}
In the framework of STEGR, the contorsion tensor coincides with the antisymmetric part of the affine connection, leading to the relation, $T^{k}_{ij}=2\Gamma^{k}_{[ij]}=\Gamma^{k}_{ij}-\Gamma^{k}_{ji}$.
$L^{k}_{ij}$ describes the deformation and is given by:
\begin{equation} 
	L^{k}_{ij}=\frac{1}{2}Q^{k}_{ij}+Q_{(_{i}~k~_{j})}. \label{eq13} 
\end{equation}
The superpotential associated with non-metricity is defined as
\begin{equation} 			            
    P^{kij}=-\frac{1}{4}Q^{kij}+\frac{1}{2}Q^{(ij)k}+\frac{1}{4}(Q^{k}-\tilde{Q}^{k})g^{ij}-\frac{1}{4}\delta^{k(_{i}Q_{j})}, \label{eq14} 
\end{equation}
where the traces of the non-metricity tensor, $Q^{k}$ and $\tilde{Q}^{k}$, are defined as
\begin{equation} 
	Q_{k}\equiv {Q_{k}}^{i}_{i}, ~~~~~~~~~~~ \tilde{Q}^{k}=Q_{i}^{ki}. \label{eq15} 
\end{equation}
Thus, the non-metricity scalar can be expressed as
\begin{equation} 
	Q=-g^{ij}\Big(L^{k}_{lj}L^{l}_{ik}-L^{l}_{il}L^{k}_{ij}\Big)=-Q_{kij}P^{kij}. \label{eq16} 
\end{equation}
To obtain the gravitational field equations, one varies the Einstein-Hilbert action in Eq.~\eqref{eq8} with respect to the metric tensor, $g_{ij}$, leading to
\begin{equation}    
	\frac{2}{\sqrt{-g}}\nabla_{k}\Big(\sqrt{-g}f_{Q}P^{k}_{ij}\Big)+\frac{1}{2}g_{ij}f+f_{Q}\Big(P_{ikl}Q^{kl}_{j}-2Q_{kli}P^{kl}_{j}\Big)=-T_{ij}, \label{eq17} 
\end{equation}
where, $f_{Q}=\frac{\partial f}{\partial Q}$, and $T_{ij}$ represents the energy-momentum tensor corresponding to the matter contribution. The energy-momentum tensor is generally defined as
\begin{equation} 
	T_{ij}=-\frac{2}{\sqrt{-g}}\frac{\delta(\sqrt{-g}\mathfrak{L_{m}})}{\delta g^{ij}}. \label{eq18} 
\end{equation}
Now, to construct a tractable set of Einstein field equations (EFE) in the framework of $f(Q)$ gravity, we consider a static spherically symmetric space-time as
\begin{equation}
	ds^2=-e^{2\nu(r)}dt^2+e^{2\lambda(r)}dr^2+r^2(d\theta^2+sin^2\theta d\phi^2). \label{eq19}
\end{equation}
Substituting Eq.~(\ref{eq19}) in Eq.~(\ref{eq18}), we obtain,
\begin{equation}
	Q=-\frac{2e^{-2\lambda(r)}}{r}\Big(2\nu'(r)+\frac{1}{r}\Big), \label{eq20}
\end{equation}
Where, the overhead prime denotes derivative with respect to radial coordinate $(r)$. Again, the isotropic perfect fluid distribution is characterised by
\begin{equation}
	T_{ij}=(\rho+p)u_{i}u_{j}+pg_{ij}, \label{eq21}
\end{equation} 
where, $u_{i}:u^{i}u_{i}=-1$ stands for the four velocity associated with the fluid, $\rho$ is the energy density and $p$ represents the isotropic pressure component. By integrating the equation of motion given in Eq. (\ref{eq17}) with the isotropic perfect fluid configuration outlined in Eq. (\ref{eq19}), we obtain a generalised set of non-vanishing components of the EFE, expressed as follows:
\begin{eqnarray}
	\frac{f(Q)}{2}-f_{Q}\Big[Q+\frac{1}{r^{2}}+\frac{2e^{-2\lambda}}{r}(\nu'+\lambda')\Big]=8\pi\rho, \label{eq22} \\
	-\frac{f(Q)}{2}+f_{Q}\Big[Q+\frac{1}{r^{2}}\Big]=8\pi p, \label{eq23} \\
	-\frac{f(Q)}{2}+f_{Q}\Big[\frac{Q}{2}-e^{-2\lambda}\Big\{\nu''+2~\Big(\frac{\nu'}{2}+\frac{1}{2r}\Big)(\nu'-\lambda')\Big\}\Big]=8\pi p, \label{eq24} \\
	\frac{cot\theta}{2}Q'f_{QQ}=0, \label{eq25}
\end{eqnarray}
where, the overhead prime denotes derivative with respect to $r$. In a straightforward way, the solution of Eq.~(\ref{eq25}) yields,
\begin{equation}
	f(Q)=\alpha_{0}+\alpha_{1}Q. \label{eq26}
\end{equation} 
Here, $\alpha_{0}$ and $\alpha_{1}$ are characteristic constants, where, $\alpha_{0}$ has dimensions of $Km^{-2}$ and $\alpha_{1}$ is dimensionless. Using Eqs.~(\ref{eq20}) and (\ref{eq26}), we can further refine the set of Eqs.~(\ref{eq22}), (\ref{eq23}), (\ref{eq24}), (\ref{eq25}) in the following form:
\begin{eqnarray}
	\frac{1}{2r^{2}}\Big[r^{2}\alpha_{0}-2\alpha_{1}e^{-2\lambda}(2r\lambda'-1)-2\alpha_{1}\Big]=8\pi\rho, \label{eq27}\\
	\frac{1}{2r^{2}}\Big[-r^{2}\alpha_{0}-2\alpha_{1}e^{-2\lambda}(2r\nu'+1)+2\alpha_{1}\Big]=8\pi p, \label{eq28} \\
	\frac{e^{-2\lambda}}{2r}\Big[-r\alpha_{0}e^{2\lambda}-2r\alpha_{1}\nu''-2\alpha_{1}(r\nu'+1)(\nu'-\lambda')\Big]=8\pi p. \label{eq29}
\end{eqnarray}  
Now, a sophisticated way to relate the geometry and matter involves employing conformal symmetry, specifically through the application of Conformal Killing Vectors (CKVs), as expressed by the following equation:
\begin{equation}
	\mathcal{L_{\eta}}g_{ij}=2\psi g_{ij}, \label{eq30}
\end{equation}
where, $\mathcal{L}$ represents the operator of Lie derivative and $\psi$ denotes the conformal factor. The vector $\eta$ generates the conformal symmetry, ensuring that the metric $g$ is conformally mapped onto itself along the direction of $\eta$. It is important to note that neither $\eta$ nor $\psi$ need to be static, even when dealing with a static metric \cite{Boehmer,Boehmer1}. From Eq.~(\ref{eq30}), we have the following three notions: (i) for $\psi=0$, one obtains an asymptotically flat space-time which in turn points toward the vanishing Weyl tensor, (ii) for $\psi=constant$, one obtains homothetic motion, and (iii) for $\psi=\psi(r,t)$, one obtains conformal vectors. Using Eq.~(\ref{eq19}) in Eq.~(\ref{eq30}), the conformal killing equations are expressed as,
\begin{equation}
	\eta^{1}\nu^{1}=\psi, ~~~~~~\eta^{1}\lambda'+\eta^{1}_{,1}=\psi,~~~~~\eta^{1}=\psi, r~~~~~\eta^{0}=c_{const}. \label{eq31}
\end{equation} 
Here, the overhead prime and comma represent, respectively, the normal and partial derivatives with respect to r and $c_{const}$ is a constant. Considering the spherically symmetric line element, expressed in Eq.~(\ref{eq19}), Eq.~(\ref{eq31}) yields,
\begin{equation}
	e^{2\nu}=c_{1}^{2}r^{2}, \label{eq32}
\end{equation}
\begin{equation}
	e^{2\lambda}=\Bigg(\frac{c_{2}}{\psi}\Bigg)^{2}, \label{eq33}
\end{equation}
and, 
\begin{equation}
	\eta^{i}=c_{const}~\delta_{0}^{i}+(\psi r)~\delta_{1}^{i}, \label{eq34}
\end{equation}
where, $c_{1}$ and $c_{2}$ are constants of integration and $\delta$ represents the Kronecker delta function.

In view of the conformal Killing equations, expressed in Eqs.~(\ref{eq32}) and (\ref{eq33}), the modified form of the EFE in the framework of $f(Q)$ gravity, considering conformal symmetry, takes the following form:
\begin{eqnarray}
	\frac{\alpha_{0}}{2}-\frac{\alpha_{1}}{r^{2}}+\frac{\alpha_{1}\psi(\psi+2r\psi')}{c_{2}^{2}r^{2}}=8\pi\rho, \label{eq35}\\
	-\frac{\alpha_{0}}{2}+\frac{\alpha_{1}}{r^{2}}-\frac{3\alpha_{1}\psi^{2}}{c_{2}^{2}r^{2}}=8\pi p, \label{eq36} \\
	-\frac{\alpha_{0}}{2}-\frac{\alpha_{1}\psi(\psi+2r\psi')}{c_{2}^{2}r^{2}}=8\pi p. \label{eq37}
\end{eqnarray}
In the upcoming sections, we will solve this above set of equations for the different layers of gravastar-like structures and study their physical properties. 
\section{Construction of gravastar-like structures with SIQM shell}\label{sec4}
\begin{figure}[h!]
	\centering
	\includegraphics[width=8cm]{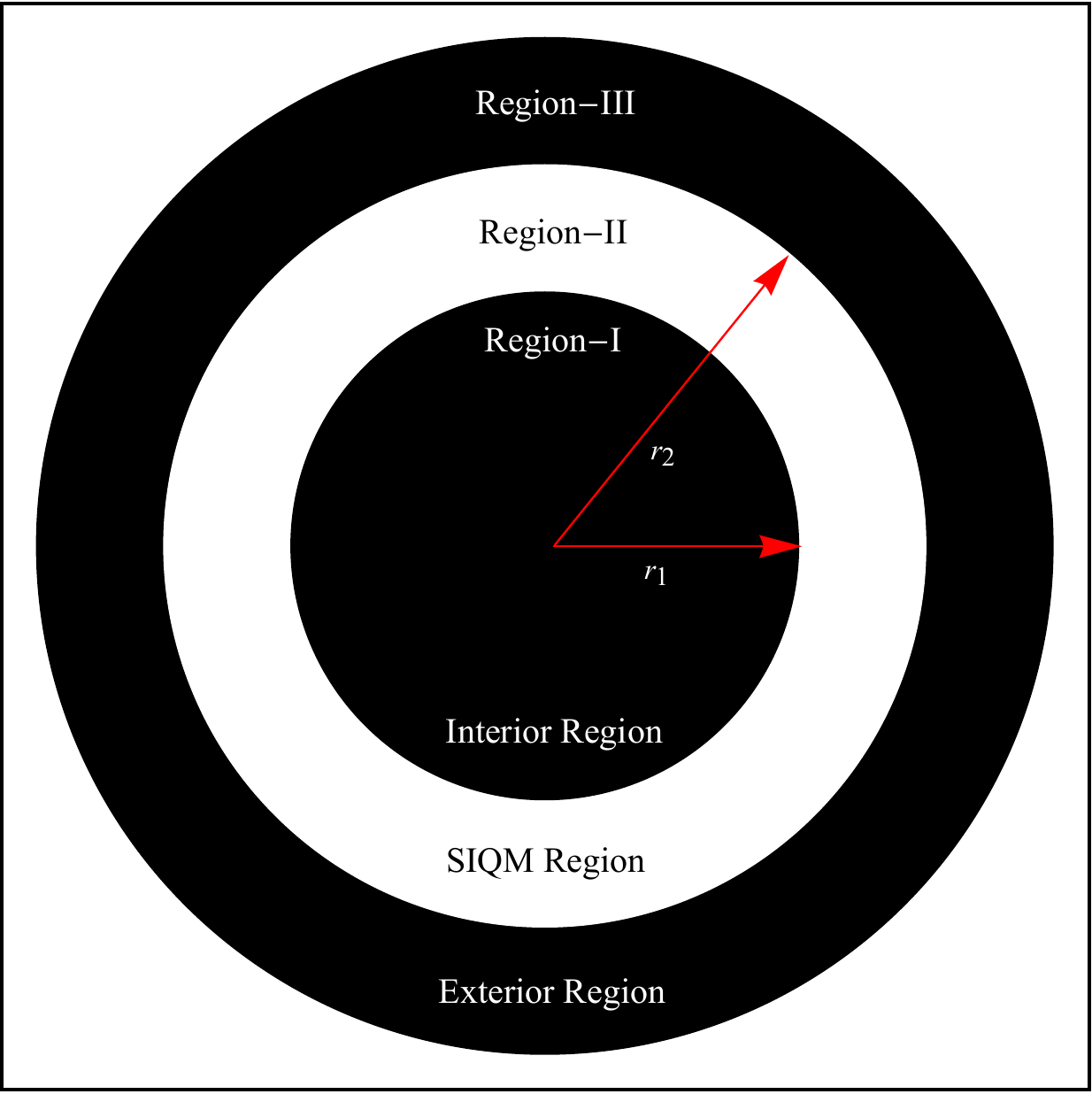}
	\caption{Schematic diagram of gravastar-like structures with SIQM shell}
	\label{fig1}
\end{figure}

The present study originates from the intricate structure of gravastars. Following the Mazur and Mottola hypothesis \cite{Mazur,Mazur1,Mazur2}, we note that the interior and exterior regions of a gravastar are separated by the presence of a finite slice of thin shell. The thin shell is composed of an ultra-relativistic stiff fluid, characterised by the EoS, $p=\rho$, which was proposed by Zel'dovich \cite{Zeldovich,Zeldovich1} in the context of cold baryonic universe. Interestingly, the EoS of the thin shell pushes the boundary of the causality to its maximum limit $\Big(\frac{dp}{d\rho}=1\Big)$. We begin our investigation from this point and construct a three-layered stellar structure which is akin to a gravastar, but with the thin shell now composed of SIQM which maintains the apex of causality as expressed in Eq.~(\ref{eq7a}). In Figure~\ref{fig1}, we illustrate the schematic diagram of gravastar-like structures with SIQM shell.    
\subsection{Interior region:} The interior region of a gravastar is hypothesised as a de-Sitter space-time to remove the problem of central singularity persistent in the context of a classical BH \cite{Mazur,Mazur1,Mazur2}. The characterisation of such a space-time is based on the particular choice of EoS of the form, $p=-\rho$. In this region, the presence of negative or repulsive pressure generates a radially outward force at every point on the shell, effectively counteracting the inward pull of gravitational collapse. Simultaneously, the negative energy density within this region corresponds to a positive cosmological constant. This specific EoS is often referred to as a "degenerate vacuum" or "$\rho$-vacuum," serving as a fundamental example of a dark energy EoS. Now, using the EoS in the set of Eqs.~(\ref{eq35}), (\ref{eq36}), (\ref{eq37}), we obtain a differential equation of the form:
\begin{equation}
	\frac{2\alpha_{1}\psi\Big(r\psi'-\psi\Big)}{c_{2}^{2}r^{2}}=0. \label{eq38}
\end{equation}
There are two possible solutions of the above equation: (i) $\psi=0$ and (ii) $\psi=rc_{3}$, where $c_{3}$ is an integration constant. Notably, we discard $\psi=0$ as it describes an asymptotically flat solution and adhere to $\psi=rc_{3}$, to study the interior properties of gravastars in this framework. Now, the metric potentials for the interior region can be expressed in terms of conformal parameter $(\psi)$ in the following form:
\begin{equation}
	e^{-2\lambda}=\Bigg(\frac{\psi}{c_{2}}\Bigg)^{2}=r^{2}\Bigg(\frac{c_{3}}{c_{2}}\Bigg)^{2}=c_{4}^{2}r^{2}, \label{eq39}
\end{equation}  
and 
\begin{equation}
	e^{2\nu}=c_{1}^{2}r^{2}, \label{eq40}
\end{equation}
where, $c_{4}=\frac{c_{3}}{c_{2}}$ is a constant. It is interesting to note that, the $g_{rr}$ and $g_{tt}$ metric components of the interior region are inversely and directly proportional to $r^{2}$, respectively. Further, the energy density and pressure associated to the interior region can be expressed as
\begin{equation}
	\rho=\frac{c_{2}^{2}r^{2}\alpha_{0}-2c_{2}^{2}\alpha_{1}+6c_{3}^{2}r^{2}\alpha_{1}}{16\pi c_{2}^{2}r^{2}}=-p. \label{eq41}
\end{equation} 
Additionally, the positive energy density of the interior region $(\rho>0)$ restricts the parameter space in the following way: for $r>0$, $\alpha_{1}<0$, $\alpha_{0}>\frac{2\alpha_{1}}{r^{2}}$ and $c_{2}>0$, we obtain, $c_{3}<\sqrt{\frac{2c_{2}^{2}\alpha_{1}-c_{2}^{2}r^{2}\alpha_{0}}{6r^{2}\alpha_{1}}}$. We note that, the energy density $(\rho)$ and isotropic pressure $(p)$ undergo central singularity, which is natural in stellar models with CKVs. 
\subsection{Thin shell with SIQM:} Using Eq.~(\ref{eq7a}) in Eqs.~(\ref{eq35}), (\ref{eq36}), (\ref{eq37}), we obtain the ordinary differential equation in the form:
\begin{equation}
	\frac{2\alpha_{1}\psi\psi'}{c_{2}^{2}r}+\frac{4\alpha_{1}\psi^{2}}{c_{2}^{2}r^{2}}-\frac{2\alpha_{1}}{r^{2}}+\alpha_{0}-16\pi B_{g}=0. \label{eq42}
\end{equation}
Solution of the Eq.~(\ref{eq42}) leads to the form of conformal parameter $(\psi)$ as
\begin{equation}
	\psi=\frac{\sqrt{16\pi B_{g}c_{2}^{2}r^{6}-c_{2}^{2}r^{6}\alpha_{0}+3c_{2}^{2}r^{4}\alpha_{1}+6\alpha_{1}c_{5}}}{\sqrt{6\alpha_{1}}r^{2}}. \label{eq43}
\end{equation}
Substituting Eq.~(\ref{eq43}) in Eqs.~(\ref{eq32}) and (\ref{eq33}), we obtain
\begin{equation}
	e^{-2\lambda}=\frac{1}{2}-\frac{r^{2}\alpha_{0}}{6\alpha_{1}}+\frac{8\pi B_{g}r^{2}}{3\alpha_{1}}+\frac{c_{5}}{c_{2}^{2}r^{4}},  \label{eq44}
\end{equation}
and 
\begin{equation}
	e^{2\nu}=c_{1}^{2}r^{2}. \label{eq45}
\end{equation}
Further, within this parameter space, utilising Eq.~(\ref{eq43}) in Eqs.~(\ref{eq35}), (\ref{eq36}), (\ref{eq37}), we obtain expression for SIQM pressure and energy density on the thin shell as
\begin{equation}
	\rho=B_{g}-\frac{\alpha_{1}(6c_{5}+c_{2}^{2}r^{4})}{16\pi c_{2}^{2}r^{6}}, \label{eq46}
\end{equation}
and 
\begin{equation}
	p=-B_{g}-\frac{\alpha_{1}(6c_{5}+c_{2}^{2}r^{4})}{16\pi c_{2}^{2}r^{6}}. \label{eq47}
\end{equation}
\subsection{Exterior region:} To establish an exterior solution in the framework of linear $f(Q)$ action, we follow the procedure laid out by Wang et al. \cite{Wang}. Now, the off-diagonal component described in Eq.~(\ref{eq21}), the solutions within the $f(Q)$ gravity framework are constrained to two specific cases:
\begin{eqnarray} 
	f_{QQ}=0, \Rightarrow f(Q)=\alpha_{0} + \alpha_{1}Q, \label{eq48} \\
	Q'=0, \Rightarrow Q=Q_{0}, \label{eq49} 
\end{eqnarray}
where $\alpha_{0}$, $\alpha_{0}$ are defined earlier, and $Q_{0}$ is a constant. These solutions mirror those derived in $f(T)$ gravity, as discussed by B\"ohmer et al. \cite{Bohmer}, where similar constraints $f_{TT} = 0$ or $T' = T_0$ were used for static, spherically symmetric configurations under a diagonal tetrad framework. Furthermore, if the cosmological constant is defined as $\frac{\alpha_{0}}{\alpha_{1}}$, the first solution in equation \eqref{eq48} reduces to GR, aligning with symmetric teleparallel gravity. Further, for a vacuum scenario, where $\rho$ and $p$ vanish, the equations of motion (eqs. \eqref{eq20}, \eqref{eq22}, \eqref{eq23}, \eqref{eq24}) simplify under the assumption of Eq.~(\ref{eq48}) to the following form:
\begin{equation} 
	\lambda'(r)+\nu'(r)=0, \label{eq50} 
\end{equation} 
\begin{equation} 
	Q=\frac{\alpha_{0}}{\alpha_{1}}-\frac{2}{r^{2}}. \label{eq51} 
\end{equation}
From Eq.~(\ref{eq50}), we deduce:
\begin{equation} 
	\nu(r)=-\lambda(r)+\lambda_{0}, \label{eq52} 
\end{equation}
where $\lambda_{0}$ is an integration constant that can be absorbed into the solution by rescaling the time coordinate, $t\rightarrow e^{-\lambda_{0}}t$. Consequently, the $g_{rr}$ and $g_{tt}$ components become reciprocally related, as noted by Wang et al. \cite{Wang}. This simplifies to:
\begin{equation} 
	\nu(r)=-\lambda(r). \label{eq53} 
\end{equation}
Using Eqs.~(\ref{eq20}), \eqref{eq51}, and \eqref{eq52}, we derive the exterior $g_{rr}$ component:
\begin{equation} 
	e^{2\lambda}=\Bigg(1-\frac{c_{6}}{r}-\frac{\alpha_{0}}{6\alpha_{1}}r^{2}\Bigg)^{-1}, \label{eq54}
\end{equation}
where $c_{6}$ is an integration constant, chosen for convenience. Similarly, using eqs. \eqref{eq53} and \eqref{eq54}, the $g_{tt}$ component is obtained as
\begin{equation} 
	e^{2\nu}=\Bigg(1-\frac{c_{6}}{r}-\frac{\alpha_{0}}{6\alpha_{1}}r^{2}\Bigg). \label{eq55} 
\end{equation}
Thus, the line element for the spherically symmetric vacuum exterior spacetime in $f(Q)$ gravity, describing a compact stellar object, is:
\begin{equation} 
	ds^{2}=\Bigg(1-\frac{c_{6}}{r}-\frac{\alpha_{0}}{6\alpha_{1}}r^{2}\Bigg)dt^{2}+\Bigg(1-\frac{c_{6}}{r}-\frac{\alpha_{0}}{6\alpha_{1}}r^{2}\Bigg)^{-1}dr^{2}+r^{2}(d\theta^{2}+\sin^{2}\theta d\phi^{2}). \label{eq56} 
\end{equation}
Upon examining this metric, we observe three key features: (i) when $c_{6}=2M$, and (ii) when the cosmological constant is $\Lambda=\frac{\alpha_{0}}{\alpha_{1}}$, the metric simplifies to the Schwarzschild-(anti) de Sitter solution:
\begin{equation}
	 ds^{2}=\Bigg(1-\frac{2M}{r}-\frac{1}{3}\Lambda r^{2}\Bigg)dt^{2}+\Bigg(1-\frac{2M}{r}-\frac{1}{3}\Lambda r^{2}\Bigg)^{-1}dr^{2}+r^{2}(d\theta^{2}+\sin^{2}\theta d\phi^{2}), \label{eq57} 
\end{equation}
where $M$ is the total mass of the stellar object with radius $R$. In the absence of the cosmological constant, the metric further reduces to the Schwarzschild solution in the form:
\begin{equation} 
	ds^{2}=-\Bigg(1-\frac{2M}{r}\Bigg)dt^{2}+\Bigg(1-\frac{2M}{r}\Bigg)^{-1}dr^{2}+r^{2}(d\theta^{2}+\sin^{2}\theta d\phi^{2}). \label{eq58} 
\end{equation}
Therefore, the close relationship between Eqs.~(\ref{eq56}) and (\ref{eq57}) suggests that such solutions are attainable exclusively under a linear form of $f(Q)$ gravity. 
\section{Junction condition}\label{sec5} 
A gravastar consists of three distinct regions: an interior core, a thin shell in the intermediate region, and an exterior spacetime. The thin shell acts as a boundary, smoothly connecting the interior and exterior geometries. The fundamental junction condition ensuring a seamless transition at the hypersurface $(r=R)$ is governed by the Israel formalism \cite{Israel,Israel1}. It is important to note that while the metric coefficients remain continuous across the interface, their derivatives may exhibit discontinuities. To analyze the surface stresses at the junction, we employ the Darmois-Israel junction conditions \cite{Israel}-\cite{Darmois}. Within Einstein gravity, the intrinsic surface stress-energy tensor $S_{\gamma\beta}$ is given by the Lanczos equation \cite{Lanczos}-\cite{Musgrave} in coordinates $X^{\alpha}=(t,r,\theta,\phi)$ as follows:
\begin{equation} 
	{S}^{\gamma}_{\beta}=-\frac{1}{8\pi} \Big[{K}^{\gamma}_{\beta}-{\delta}^{\gamma}_{\beta}{K}^{\kappa}_{\kappa}\Big], \label{eq59} 
\end{equation}
where, ${K}_{\gamma\beta}={K}^{+}_{\gamma\beta}-{K}^{-}_{\gamma\beta}$, with the superscripts $(+)$ and $(-)$ denoting values at the exterior and interior interfaces, respectively. The expression for the second fundamental form is
\begin{equation}
	{K}_{\gamma\beta}^{\pm}=-{\eta}_{\tau}^{\pm} \Big(\frac{\partial^2 {X_{\tau}}}{\partial{\zeta^{\gamma}} \partial{\zeta^{\beta}}}+{\Gamma}_{\alpha\beta}^{\tau} \frac{\partial{X_{\alpha}}}{\partial {\zeta^{\gamma}}} \frac{\partial{X_{\beta}}}{\partial{\zeta^{\beta}}}\Big). \label{eq60}
\end{equation}
The double-sided normal to the hypersurface is expressed as
\begin{equation}
	\eta_{\tau}^{\pm}={\pm} \Big|{g}^{\alpha\beta} \frac{\partial{f}}{\partial{x^{\alpha}}} \frac{\partial{f}}{\partial{x^{\beta}}}\Big|^{-\frac{1}{2}}\frac{\partial{f(r)}}{\partial{x^{\tau}}}, \label{eq61}
\end{equation}
with $\eta^{\tau}\eta_{\tau}=1$. By applying the Lanczos equation \cite{Lanczos}-\cite{Musgrave}, the surface stress-energy tensor at the junction interface takes the form, ${S}_{\gamma\beta}=diag(-\Sigma, \mathfrak{P}, \mathfrak{P}, \mathfrak{P} )$, where $\Sigma$ and $\mathfrak{P}$ correspond to the surface energy density and surface pressure, respectively, given by
\begin{eqnarray}
	\Sigma=-\frac{1}{4\pi R}{\Big(\sqrt{f(r)}\Big)}^{+}_{-}~, \label{eq62} \\
	\mathfrak{P}=-\frac{\Sigma}{2}+\frac{1}{16\pi} \Big(\frac{f'(r)}{\sqrt{f(r)}}\Big)^{+}_{-}, \label{eq63}
\end{eqnarray}
where, $f(r)^{+}_{-}$ denotes the metric components in the exterior and interior regions, respectively. Using Eqs.(\ref{eq58}) and (\ref{eq39}), the expressions for surface energy density $\Sigma$ reduces to
\begin{equation}
	\Sigma=\frac{\sqrt{c_{4}^{2}R^{2}}-\sqrt{1-\frac{2M}{R}}}{4\pi R}. \label{eq64}
\end{equation}
Using Eq.~(\ref{eq64}), the mass of the gravastar shell is expressed as
\begin{equation}
	M_{shell}=4\pi R^{2}\Sigma=R\Bigg(\sqrt{c_{4}^{2}R^{2}}-\sqrt{1-\frac{2M}{R}}\Bigg). \label{eq65}
\end{equation}
\section{Boundary conditions}\label{sec6}
The physical features of the present model depend on the characteristic constants appearing in this framework. To determine these constants, we match the interior and thin shell solutions along with thin shell and exterior solutions at interior radius $(r=r_{1})$ and exterior radius $(r=r_{2})$, respectively in the following way:
\begin{equation}
	c_{4}^{2}r_{1}^{2}=\frac{1}{2}-\frac{r_{1}^{2}\alpha_{0}}{6\alpha_{1}}+\frac{8\pi B_{g}r_{1}^{2}}{3\alpha_{1}}+\frac{c_{5}}{c_{2}^{2}r_{1}^{4}}, \label{eq66}
\end{equation}  
\begin{equation}
	\frac{1}{2}-\frac{r_{2}^{2}\alpha_{0}}{6\alpha_{1}}+\frac{8\pi B_{g}r_{2}^{2}}{3\alpha_{1}}+\frac{c_{5}}{c_{2}^{2}r_{2}^{4}}=1-\frac{2M}{r_{2}}, \label{eq67}
\end{equation}  
and 
\begin{equation}
	c_{1}^{2}r_{2}^{2}=1-\frac{2M}{r_{2}}. \label{eq68}
\end{equation}
Now, considering the requirement that the Buchdahl limit, i.e., $\frac{2M}{r}<\frac{8}{9}$, \cite{Buchdahl} must be maintained for a physically acceptable stellar model, we have considered the total mass $(M)$ to be $2.5~M_{\odot}$ and three characteristic radii for the present analysis, {\it viz.}, 9-9.009 Km, 10-10.009 Km and 11-11.009 Km \cite{Bhattacharjee}. Moreover, based on the recent work of Bhattacharjee and Chattopadhyay \cite{Bhattacharjee1}, we have set $\alpha_{0}=10^{-46}~Km^{-2}$ and $\alpha_{1}=-0.5$. Now, the positivity of $M_{shell}$, i.e., $M_{shell}>0$, yields, $c_{4}>\sqrt{\frac{R-2M}{R^{3}}}$, while the stipulation that the surface redshift $(Z_{s})$ of thin shell of an isotropic configuration is less than 2, i.e., $Z_{s}<2$ \cite{Buchdahl}, produces, $c_{4}<\sqrt{\frac{97}{81R^{2}}-\frac{2M}{R^{3}}+\frac{8}{9}\sqrt{\frac{R-2M}{R^{5}}}}$. The upper and lower bounds of $c_{4}$ are tabulated in Table~\ref{tab1}. Now, the range of $B_{g}$ associate with stable SQM, relative to neutron, is $57.55-95.11~MeV/fm^{3}$ \cite{Madsen}. Now, for a comprehensive analysis and determination of constants, we have considered two combinations here, {\it viz.}, (i) $B_{g}=70~MeV/fm^{3}$ with different characteristic radii and (ii) $R=10-10.009~Km$ with $B_{g}$ ranging from $57.55-95.11~MeV/fm^{3}$. It must be noted that the arbitrary choice of $B_{g}=70~MeV/fm^{3}$ has been further supported by the work of Aziz et al. \cite{Aziz}. The results for the two scenarios are tabulated in Tables~\ref{tab2} and \ref{tab3}, respectively. Notably, we have considered $c_{5}=0.0001$ without the loss of any generality. 
\begin{table}[h!]
	\centering
	\caption{Bounds on $c_{4}$.}
	\label{tab1}
	\begin{tabular}[htbp]{@{}cc@{}}
		\hline
		Radius (Km) & Bounds \\
		\hline
		9-9.009 & $0.047<c_{4}<0.096$ \\
		10-10.009 & $0.051<c_{4}<0.095$ \\
		11-11.009 & $0.052<c_{4}<0.092$ \\
		\hline
	\end{tabular}
\end{table}
\begin{table}[h!]
	\centering
	\caption{Numerical determination of necessary constants for $B_{g}=70~MeV/fm^{3}$ and different characteristic radii.}
	\label{tab2}
 	\begin{tabular}[htbp]{@{}cccccc@{}}
 	    \hline
 	    \multirow{2}{*}{M $(M_{\odot})$} & \multicolumn{2}{c}{Radius (Km)} & \multirow{2}{*}{$c_{1}$} & \multirow{2}{*}{$c_{2}$} & \multirow{2}{*}{$c_{5}$} \\
 	    & $r_{1}$ & $r_{2}$ &&&\\  
 	    \hline
 	    \multirow{3}{*}{$2.5~M_{\odot}$} & 9 & 9.009 & 0.047 & 0.00032 & \multirow{3}{*}{0.0001} \\
 	    & 10 & 10.009 & 0.051 & 0.00018 & \\
 	    & 11 & 11.009 & 0.052 & 0.00012 & \\
 	    \hline
 	  \end{tabular}
 \end{table}
\begin{table}[ht!]
	\centering
	\caption{Numerical determination of necessary constants for different $B_{g}$ and $R=10-10.009~Km$.}
	\label{tab3}
	\begin{tabular}[htbp]{@{}cccc@{}}
		\hline
		$B_{g}~(MeV/fm^{3})$ & $c_{1}$ & $c_{2}$ & $c_{5}$ \\
		\hline
		57.55 & \multirow{3}{*}{0.051} & 0.00019 & \multirow{3}{*}{0.0001} \\
		70.00 & & 0.00018 &  \\
		95.11 & & 0.00016 &  \\
		\hline
	\end{tabular}
\end{table} 
Additionally, following Table~\ref{tab1}, we have considered $c_{4}=0.08$ for the physical analysis.
\section{Mass of the thin shell} \label{sec7}  In the present context, in view of Eq.~(\ref{eq65}), we note that the mass of the thin shell is independent of the matter distribution of the shell region. Furthermore, it must be noted that within the bounds described in Table~\ref{tab1}, we have set $c_{4}=0.08$. Hence, for all the characteristic radii, $c_{4}$ remains constant and following Eq.~(\ref{eq65}), it is evident that $M_{shell}$ becomes solely radius dependent. Now, using the numerical values of Table~\ref{tab2} in Eq.~(\ref{eq65}), we have obtained the mass of the thin shell and the results are tabulated in Table~\ref{tab4}.
\begin{table}[ht!]
	\centering
	\caption{Mass of the thin shell $(M_{shell})$ different characteristic radii.}
	\label{tab4}
	\begin{tabular}[htbp]{@{}cc@{}}
		\hline
		Radius (Km) & $M_{shell}~(M_{\odot})$ \\
		\hline
		9.009 & 1.80 \\
		10.009 & 1.95 \\
		11.009 & 2.28 \\
		\hline
	\end{tabular}
\end{table} 
From Table~\ref{tab4}, we note that with increasing radius, $M_{shell}$ increases. 
\section{Physical features of the proposed model}\label{sec8}
The key attributes associated with gravastars are the proper length, energy and entropy of the thin shell. In this section, we have utilised Tables~\ref{tab2} and \ref{tab3} and investigated these features within the parameter space used here. 
\subsection{Proper Length:} 
\begin{figure}[h!]
	\centering
	\begin{subfigure}[t]{0.45\textwidth}
		\centering
		\includegraphics[width=1\textwidth]{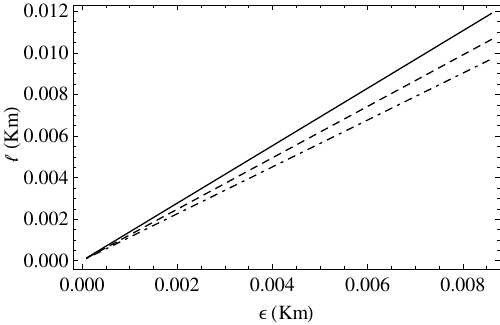}
		\caption{}
		\label{fig2a}
	\end{subfigure}
	\hfill
	\begin{subfigure}[t]{0.45\textwidth}
		\centering
		\includegraphics[width=1\textwidth]{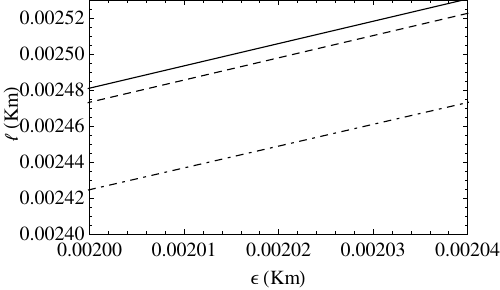}
		\caption{}
		\label{fig2b}
	\end{subfigure}
	\caption{Variation of proper length $(\ell)$ with shell thickness $(\epsilon)$ for (a) different characteristic radii and $B_{g}=70~MeV/fm^{3}$. Here, the solid, dashed and dotdashed lines represent 9-9.009, 10-10.009 and 11-11.009 Km, respectively, (b) for different bag constant $(B_{g})$ and radius 10-10.009 Km. Here, the solid, dashed and dotdashed lines represent $B_{g}=57.55,~70$ and $95.11~MeV/fm^{3}$, respectively.}
	\label{fig2}
\end{figure}

One of the characteristic features of a gravastar is the proper length of the thin shell. Using Eq.~(\ref{eq44}), the proper length is analytically expressed as
\begin{equation}
	\ell=\int_{r_{1}}^{r_{2}} e^{\lambda}~dr=\int_{r_{1}}^{r_{2}}\frac{dr}{\sqrt{\frac{1}{2}-\frac{r_{1}^{2}\alpha_{0}}{6\alpha_{1}}+\frac{8\pi B_{g}r_{1}^{2}}{3\alpha_{1}}+\frac{c_{5}}{c_{2}^{2}r_{1}^{4}}}}. \label{eq69}
\end{equation}  
To avoid the mathematical complexities and present a more substantial verification, we have graphically illustrated the variation of proper length with respect to the thickness of the shell $(\epsilon=r_{2}-r_{1})$ in Figure~\ref{fig2}. 
From Figure~\ref{fig2a}, we note that for a particular radius, the proper length increases with increasing shell thickness. Interestingly, we note that with increasing radii, the proper length decreases, which may be described as follows: the thickness of the shell and the core region are intertwined in describing the structure and stability of a gravastar. The shell region must withstand the outward pressure of the exotic gravastar core to maintain stability. Now, with increasing radius, the core region grows in size and the core-shell distance increases. Consequently, the efforts of the interior exotic matter to prevent gravitational collapse becomes less dependent on the shell thickness. Further, with increasing radius, the curvature and gravitational potential near the shell region become less extreme. Reduced curvature near the shell at larger radii can lead to a shorter proper length because weaker gravitational gradients result in less spatial stretching. Since, the interior and shell regions are closely tied, the increasing core may result in a decreasing shell thickness, and hence decreasing proper length, while still maintaining stability. On the other hand, from Figure~\ref{fig2b}, we note that, for a particular radius, with increasing $B_{g}$, the proper length decreases, which may be attributed to the fact that with increasing $B_{g}$, the difference between perturbative and non-perturbative vacuum increases, which results in an increased energy density of quarks. As a results, the strength of the interacting increases. Hence, the proper length decreases.  
\subsection{Energy:} 
\begin{figure}[h!]
	\centering
	\begin{subfigure}[t]{0.45\textwidth}
		\centering
		\includegraphics[width=1\textwidth]{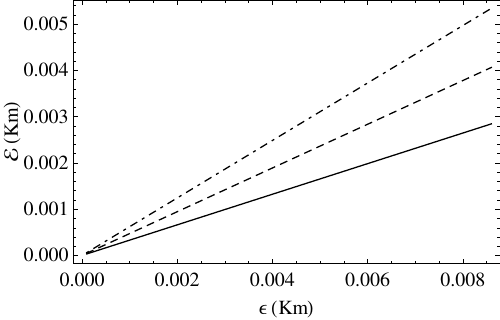}
		\caption{}
		\label{fig3a}
	\end{subfigure}
	\hfill
	\begin{subfigure}[t]{0.45\textwidth}
		\centering
		\includegraphics[width=1\textwidth]{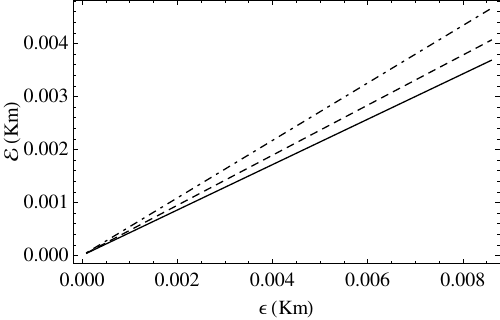}
		\caption{}
		\label{fig3b}
	\end{subfigure}
	\caption{Variation of shell energy $(\mathcal{E})$ with shell thickness $(\epsilon)$ for (a) different characteristic radii and $B_{g}=70~MeV/fm^{3}$. Here, the solid, dashed and dotdashed lines represent 9-9.009, 10-10.009 and 11-11.009 Km, respectively, (b) for different bag constant $(B_{g})$ and radius 10-10.009 Km. Here, the solid, dashed and dotdashed lines represent $B_{g}=57.55,~70$ and $95.11~MeV/fm^{3}$, respectively.}
	\label{fig3}
\end{figure}

In the present scenario, the shell region is characterised by SIQM EoS and the energy of the shell is expressed as
\begin{equation}
	\mathcal{E}=\int_{r_{1}}^{r_{2}} 4\pi r^2\rho~dr. \label{eq70}
\end{equation}
Substituting the energy density of the thin shell, as expressed in Eq.~(\ref{eq46}), in Eq.~(\ref{eq70}), we get the energy contained within the thin shell. However, we have opted for the graphical representation of the result and it is shown in Figure~\ref{fig3}. 
Figure~\ref{fig3a} illustrates that with increasing shell thickness, the energy increases. Hence, more energetic fluid can be found toward the outer edge of the shell. However, since, with increasing radius the proper length decreases, the fluid inside the shell gets confined in more tighter space. As a result, the energy increases. Moreover, as the magnitude of strong interaction increases with increasing $B_{g}$, the energy of the shell increases as demonstrated in Figure~\ref{fig3b}.  
\subsection{Entropy:} Entropy quantifies the degree of disorder in mechanical systems. In the Mazur-Mottola model \cite{Mazur,Mazur1,Mazur2}, the vacuum interior is assumed to have zero entropy, implying that all entropy contributions originate from the thin shell. The entropy of the system is determined based on an entropy function of the given form:
\begin{equation}
	S=4\pi\int_{r_{1}}^{r_{2}} \mathfrak{s}(r)r^2e^{\alpha} dr. \label{eq71}
\end{equation}
Here, the entropy density is gives by, $\mathfrak{s}(r)=\frac{\alpha^2k_{B}^2T(r)}{4\pi\hbar^2}=\frac{\alpha k_B}{\hbar}\sqrt{\frac{p}{2\pi}}$, where $T(r)$ represents the local specific temperature. $\alpha$ is a dimensionless constant, which can be set to unity without any loss of generality. $k_{B}$ and $\hbar$ are the Boltzmann constant and reduced Planck constant, with $\hbar=\frac{h}{2\pi}$, respectively. To determine the entropy, we have replaced the pressure component of the thin shell, as expressed in Eq.~(\ref{eq47}), into Eq.~(\ref{eq71}) and similar to the other results, we have chosen to demonstrate our result through plots. 
\begin{figure}[ht!]
	\centering
	\begin{subfigure}[t]{0.45\textwidth}
		\centering
		\includegraphics[width=1\textwidth]{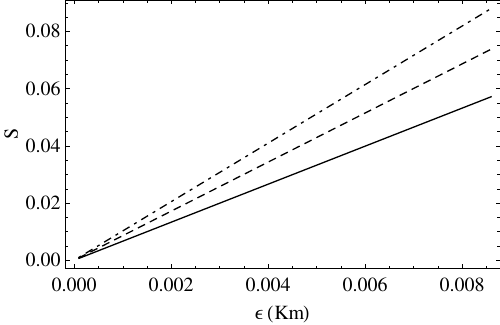}
		\caption{}
		\label{fig4a}
	\end{subfigure}
	\hfill
	\begin{subfigure}[t]{0.45\textwidth}
		\centering
		\includegraphics[width=1\textwidth]{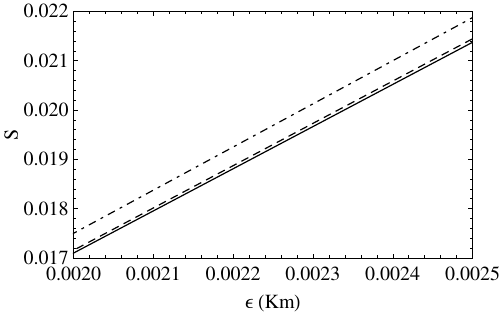}
		\caption{}
		\label{fig4b}
	\end{subfigure}
	\caption{Variation of shell entropy $(S)$ with shell thickness $(\epsilon)$ for (a) different characteristic radii and $B_{g}=70~MeV/fm^{3}$. Here, the solid, dashed and dotdashed lines represent 9-9.009, 10-10.009 and 11-11.009 Km, respectively, (b) for different bag constant $(B_{g})$ and radius 10-10.009 Km. Here, the solid, dashed and dotdashed lines represent $B_{g}=57.55,~70$ and $95.11~MeV/fm^{3}$, respectively.}
	\label{fig4}
\end{figure}
From Figure~\ref{fig4}, it is evident that with increasing shell thickness the entropy increases in both the cases. Further, as the radius and $B_{g}$ increases, the energy of the shell increases which effectively increases the number of accessible microstates. Since, entropy is strictly confined to the thin shell \cite{Mazur,Mazur1,Mazur2}, any increment in energy directly influences the entropy density, thereby contributing to the overall thermodynamic evolution of the structure. 
\section{Stability analysis} \label{sec9} The stability characteristics of the proposed gravastar model, in presence of SIQM shell, is investigated through the distinct analytical techniques of surface redshift behaviour.
\subsection{Surface redshift:} 
\begin{figure}[h!]
	\begin{subfigure}{.3\textwidth}
		\centering
		\includegraphics[width=\textwidth]{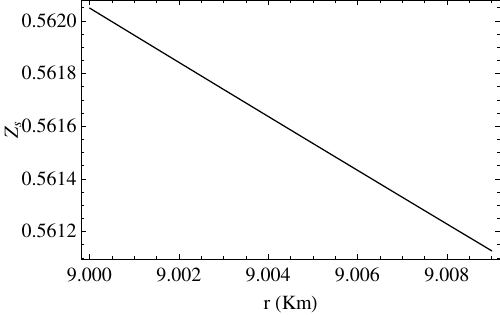}
		\caption{}
		\label{fig5a}
	\end{subfigure}%
	\hfill
	\begin{subfigure}{.3\textwidth}
		\includegraphics[width=\textwidth]{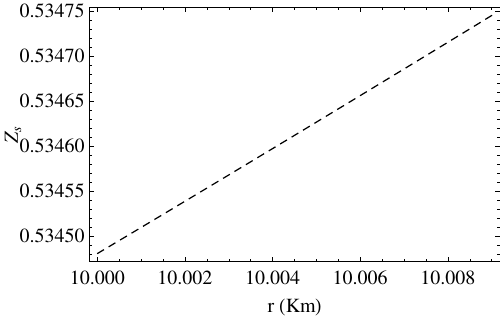}
		\caption{}
		\label{fig5b}
	\end{subfigure}
	\hfill
	\begin{subfigure}{.3\textwidth}
		\includegraphics[width=\textwidth]{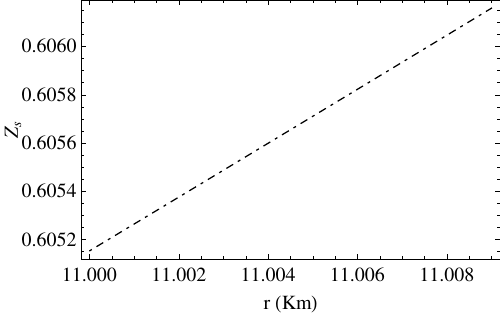}
		\caption{}
		\label{fig5c}
	\end{subfigure}
	\caption{Radial variation of surface redshift for different characteristic radii and $B_{g}=70~MeV/fm^{3}$. Here, the solid, dashed and dotdashed lines represent 9-9.009, 10-10.009 and 11-11.009 Km, respectively.}
	\label{fig5}
\end{figure}

The investigation of gravitational surface redshift plays a crucial role in assessing the stability of gravastars. This analysis is particularly significant in providing insights into their potential observational signatures. The surface redshift is defined as $Z_{s}=\frac{\lambda_{e}-\lambda_{o}}{\lambda_{e}}$ where $\lambda_{e}$ and $\lambda_{o}$ denote the emitted and observed wavelengths, respectively. According to Buchdahl, for a static, isotropic perfect fluid distribution, the surface redshift must satisfy the upper bound, $Z_{s}<2$ \cite{Buchdahl,Straumann,Bohmer}. Barraco and Hamity \cite{Barraco} further established that this restriction remains valid even in the absence of a cosmological constant. However, B\"ohmer and Harko \cite{Bohmer} demonstrated that when anisotropy is introduced in the stellar configuration, the maximum surface redshift can be extended to $Z_{s}\leq5$. Ivanov \cite{Ivanov} expanded this range further, suggesting that for anisotropic stars, the surface redshift can vary from $Z_{s}<3.84$ to $Z_{s}\leq5.211$. In a study on gravastar stability under axial perturbations, DeBenedictis et al. \cite{DeBenedictis} analysed the surface redshift parameter, concluding that their findings were in agreement with these theoretical constraints. Now, in terms of the mass of thin shell $(M_{shell})$, the surface redshift is expressed as:
\begin{equation}
	1+Z_{s}=\Bigg(1-\frac{2M_{shell}}{R}\Bigg)^{-\frac{1}{2}}. \label{eq72}
\end{equation}
Using Eq.~(\ref{eq65}) and Table~\ref{tab2}, we explore the surface redshift related to this model and the results are graphically represented in Figure~\ref{fig5}.
From Figure~\ref{fig5}, it is evident that the our model satisfies the surface redshift bound for an isotropic gravastar. Further, it must be noted that in the present scenario, for a particular radius, $c_{4}$ becomes constant. Therefore, the variation of $Z_{s}$ vs $r$ for different $B_{g}$ is not admissible here. In view of the above, we may assert that the present model is stable and physically acceptable.  
\section{Conclusion}\label{sec10}
This article is focussed on exploring gravastar like structures in the framework of $f(Q)$ gravity under conformal symmetry. Gravastars contain three layers, {\it viz.}, the interior, thin shell and the exterior region. Further, these three layers are characterised by three distinct EoS. In the present model, while preserving the characterisation of the interior and exterior regions, through the EoS, $p=-\rho$ and $p=\rho=0$, respectively, we have assumed that the thin shell is made up of SIQM, expressed through the EoS, $p=\rho-2B_{g}$ \cite{Zhang}. This choice of thin shell EoS originates from the basic structure of gravastars as introduced by Mazur and Mottola \cite{Mazur,Mazur1,Mazur2}. In that framework, the thin shell is composed of an ultra-relativistic stiff fluid, expressed through the EoS, $p=\rho$, which defines the peak of causality. Similarly, we note that the causal crest is also achieved through the SIQM EoS. Now, in the context of $f(Q)$ gravity coupled with conformal symmetry and by incorporating the SIQM shell in gravastar modeling, we have noted the following key features of the current model:
\begin{itemize}
	\item {\bf Interior region:} The fundamental distinction between a gravastar and a black hole lies in the absence of central singularity. Now, within this framework, utilising the EoS, $p=-\rho$, we have solved the system of Eqs.~(\ref{eq35}), (\ref{eq36}) and (\ref{eq37}) and obtained a differential equation as expressed in Eq.~(\ref{eq38}). Solution of Eq.~(\ref{eq38}) leads to (i) $\psi=0$ and (ii) $\psi=rc_{3}$. Now, discarding the asymptotically flat conformal solution $(\psi=0)$ and considering $\psi=rc_{3}$, we have noted singularity free solutions for the $g_{rr}$ and $g_{tt}$ components at the centre, as expressed through Eqs.~(\ref{eq39}) and (\ref{eq40}), respectively. We have also derived the analytical expression for the interior energy density $(\rho)$ and pressure $(p)$ which is written in Eq.~(\ref{eq41}). Moreover, the positive energy density $(\rho>0)$ in this region constrains the parameter space used here. 
	\item {\bf Thin shell:} Using the SIQM EoS, $p=\rho-2B_{g}$ \cite{Zhang}, we have obtained non-vanishing shell solutions in Eq.~(\ref{eq44}) and (\ref{eq45}). Further, we have obtained the expressions for the energy density $(\rho)$ and pressure $(p)$ associated with the thin shell in Eq.~(\ref{eq46}) and (\ref{eq47}). 
	\item {\bf Exterior region:} We have parametrised the vacuum $f(Q)$ solution into the Schwarzschild-(anti) de Sitter solution, as expressed in Eq.~(\ref{eq57}), which resembles Schwarzschild vacuum exterior solution, as obtained in Eq.~(\ref{eq58}), in the absence of cosmological constant. 
	\item {\bf Junction condition:} To construct a physically viable model, we enforced continuity between the interior and exterior spacetimes at the boundary hypersurface defined by $r=R$. This matching was achieved by employing the Israel junction conditions \cite{Israel,Israel1}, ensuring a smooth transition. Subsequently, the intrinsic surface energy-momentum tensor was determined. Leveraging the Lanczos equations \cite{Lanczos,Sen,Perry}, we then calculated the surface energy density, which is presented in Eq.~(\ref{eq64}). This equation has been used to define the thin shell mass $(M_{shell})$ of the gravastar, as expressed in Eq.~(\ref{eq65}).  
	\item {\bf Evaluating constant through boundary condition:} To determine the constants appearing in this framework, we have matched the interior-shell and shell-exterior solutions at interior $(r=r_{1})$ and exterior $(r=r_{2})$ radii, respectively. Further, within the stipulation of Buchdahl limit, $\frac{2M}{r}<\frac{8}{9}$ \cite{Buchdahl}, we have considered the total mass to be $2.5~M_{\odot}$ and to generate a generalised grid of models, we have considered three characteristic radii, {\it viz.}, 9-9.009 Km, 10-10.009 Km and 11-11.009 Km. Interestingly, the conditions that $M_{shell}>0$ and isotropic surface redshift, $Z_{s}<2$ provide lower and upper bounds of one of the characteristic constants, $c_{4}$, as shown in Table~\ref{tab1}. Within the limits, we have considered $c_{4}=0.08$ for the physical analysis of the present model. Moreover, the stable SQM range associated with $B_{g}$ relative to neutrons lies between $57.55-95.11~MeV/fm^{3}$ \cite{Madsen}. For a detailed numerical determination of constants, we have examined two cases: (i) $B_{g}=70~MeV/fm^{3}$ \cite{Aziz} with varying radii, and (ii) a fixed radius of $R=10-10.009~Km$ with $B_{g}$ spanning the stable range. The corresponding results are presented in Tables~\ref{tab2} and \ref{tab3}. 
	\item {\bf Mass of the thin shell:} based on Eq.~(\ref{eq65}), the mass of the thin shell is found to be independent of the matter distribution within the shell region. Notably, within the limits specified in Table~\ref{tab1}, we have fixed $c_{4}=0.08$. Consequently, $c_{4}$ remains constant for all characteristic radii, making $M_{shell}$ solely dependent on the radius, as indicated by Eq.~(\ref{eq65}). Using the numerical values from Table~\ref{tab2} in Eq.~(\ref{eq65}), we have computed the thin shell mass and found that for radii $9-9.009~Km$, $10-10.009~Km$ and $11-11.009~Km$, the mass values are $1.80,~1.95$ and $2.28~M_{\odot}$, respectively, as tabulated in Table~\ref{tab4}. Hence, with increasing radius the mass of the thin shell increases. 
	\item{\bf Key features of the model:} 
	\begin{itemize}
		\item {\bf Proper length:} Using Eq.~(\ref{eq44}), we have studied the proper length, associated with the shell region, in Eq.~(\ref{eq69}) and graphically demonstrated the results against shell thickness $(\epsilon=r_{2}-r_{1})$ in Figure~\ref{fig2}. From Figure~\ref{fig2a}, the proper length increases with shell thickness for a fixed radius but decreases with increasing radius. This suggests that the shell and core regions are interdependent in maintaining gravastar stability. As the core expands with increasing radius, the interior exotic matter's resistance to gravitational collapse relies less on the shell thickness. Additionally, larger radii reduce the curvature and gravitational potential near the shell. Less curvature implies reduced gravitational gradients, meaning reduced space-time stretching which may lead to thinner shells and shorter proper lengths while preserving structural stability. Figure~\ref{fig2b} shows that for a fixed radius, increasing $B_{g}$ reduces the proper length. This is because higher $B_{g}$ increases the difference between perturbative and non-perturbative vacua, raising the quark energy density and enhancing interaction strength, thereby reducing the proper length. 
		\item{\bf Energy:} By substituting the shell's energy density from Eq.~(\ref{eq46}) into Eq.~(\ref{eq70}), we obtain the total energy within the thin shell, represented graphically in Figure~\ref{fig3}. Figure~\ref{fig3a} shows that the shell's energy increases with thickness as well as radius, indicating that more energetic fluid is concentrated toward the shell's outer edge. Further, as the radius increases, the proper length decreases, causing the fluid to be confined in a smaller volume, thereby increasing the energy density. Additionally, Figure~\ref{fig3b} demonstrates that higher $B_{g}$ enhances the strong interaction strength, leading to increased shell energy.
		\item {\bf Entropy:} The entropy is computed by substituting the shell pressure from Eq.~(\ref{eq47}) into Eq.~(\ref{eq71}), with the results presented graphically in Figure~\ref{fig4}. The plots show that entropy increases with shell thickness in both cases. As the radius and $B_{g}$ increase, the shell's energy rises, enhancing the number of accessible microstates. Since entropy is localised within the thin shell \cite{Mazur, Mazur1, Mazur2}, any increase in energy directly raises the entropy density, contributing to the thermodynamic evolution of the gravastar structure.
	\end{itemize}
	\item{\bf Stability through surface redshift:} The stability of the model is assessed through the analysis of surface redshift $(Z_{s})$. Now, in terms of the mass of the thin shell $(M_{shell})$, $Z_{s}$ is expressed as Eq.~(\ref{eq72}). Now, within the admissible parameter space, we have shown the radial variation of $Z_{s}$ for the three characteristic radii in Figure~\ref{fig5}. From Figure~\ref{fig5}, it is evident that the condition, $Z_{s}<2$, persistent in case of an isotropic structure \cite{Buchdahl} is well maintain in this model. Hence, we may term the model stable and physically acceptable. 
\end{itemize}         



\medskip
\textbf{Acknowledgements} 
DB is thankful to Department of Science and Technology (DST), Govt. of India, for providing the fellowship vide no: DST/INSPIRE Fellowship/2021/IF210761. PKC gratefully acknowledges the support from IUCAA, Pune, India for providing the Visiting Associateship programme.

\medskip

%

\begin{thebibliography}{200}
	\bibitem{Mazur} P. O. Mazur, E. Mottola, Report number: LA-UR-01-5067 {\bf 2001}.
	\bibitem{Mazur1} P. O. Mazur, E. Mottola, Proc. Natl. Acad. Sci. USA {\bf 2004}, 101, 9545.
	\bibitem{Mazur2} P. O. Mazur, E. Mottola, Universe {\bf 2023}, 9, 88.
	\bibitem{Schwarzschild}  K. Schwarzschild, Sitzungsberichte der Koniglich Preussischen Akademie der Wissenschaften Berlin (Mathematical Physics) {\bf 1916}, pp. 189–196.
	\bibitem{Hawking} S. W. Hawking, Nature {\bf 1974}, 248, 30.
	\bibitem{Penrose} R. Penrose, Phys. Rev. Lett. {\bf 1965}, 14, 57.
	\bibitem{Ghez} A. M. Ghez et al., Astrophys. J. {\bf 1998}, 509, 678.
	\bibitem{Gillessen} S. Gillessen et al., Astrophys. J. {\bf 2009}, 692, 1075.
	\bibitem{Hawking1} S. W. Hawking, Phys. Rev. D {\bf 1976}, 13, 191.
	\bibitem{Visser} M. Visser, D. L. Wiltshire, Class. Quantum Gravity {\bf 2004}, 21, 1135.  
	\bibitem{Zeldovich} Y. B. Zel'dovich, Sov. Phys. J. Exp. Theor. Phys. {\bf 1962}, 14, 1143.	
	\bibitem{Zeldovich1} Y. B. Zel'dovich, Mon. Not. R. Astron. Soc. {\bf 1972}, 160, 1P.
	\bibitem{Chirenti} C. B. M. H. Chirenti, L. Rezzolla, Class. Quantum Gravity {\bf 2007}, 24, 4191.
	\bibitem{Carter} B. M. N. Carter, Class. Quantum Gravity {\bf 2005}, 22, 4551.
	\bibitem{Ray} S. Ray, R. Sengupta, H. Nimesh, Int. J. Mod. Phys. D {\bf 2020}, 29, 2030004.
	\bibitem{Horvat} D. Horvat, S. Ilijic, Class. Quantum Gravity {\bf 2007}, 24, 5637.
	\bibitem{DeBenedictis}  A. DeBenedictis et al., Class. Quantum Gravity {\bf 2006}, 23, 2303.
	\bibitem{Cattoen} C. Cattoen, T. Faber, M. Visser, Class. Quantum Gravity {\bf 2005}, 22, 4189. 
	\bibitem{Rahaman} F. Rahaman et al., Phys. Lett. B {\bf 2012}, 707, 319.
	\bibitem{Rahaman1} F. Rahaman et al., Int. J. Theor. Phys. {\bf 2015}, 54, 50.
	\bibitem{Ghosh}  S. Ghosh et al., Ann. Phys. (N. Y.) {\bf 2019}, 411, 167968.
	\bibitem{Bhar} P. Bhar, Astrophys. Space Sci. {\bf 2014}, 354, 457.
	\bibitem{Bilic} N. Bilic et al., J. Cosmol. Astropart. Phys. {\bf 2006}, 0602, 013.
	\bibitem{Lobo} F. Lobo, Class. Quantum Gravity {\bf 2006}, 23, 1525.
	\bibitem{Bhattacharjee} D. Bhattacharjee, P. K. Chattopadhyay, B. C. Paul, Phys. Dark Universe {\bf 2024}, 43, 101411.
	\bibitem{Bhattacharjee2} D. Bhattacharjee, P.K. Chattopadhyay, Phys. Scr. {\bf 2023}, 98, 085013.	
	\bibitem{Riess} A. G. Riess et al., Astron. J. {\bf 1998}, 116, 1009.
	\bibitem{Khlopov} M. Yu. Khlopov, B. A. Malomed, Y. B. Zel'dovich, Mon. Not. R. Astron. Soc. {\bf 1985}, 215, 575. 
	\bibitem{Khlopov1} R. V. Konoplich et al., Phys. Atom. Nucl. {\bf 1999}, 62, 1593
	\bibitem{Khlopov2} I. Dymnikova, M.Khlopov, Int. J. Mod. Phys. D {\bf 2015}, 24, 1545002.
	\bibitem{Riess1} A. G Riess et al., Astrophys. J. {\bf 2007}, 659, 98.
	\bibitem{Perlmutter} S. Perlmutter et al., Astrophys. J. {\bf 1999}, 517, 565.
	\bibitem{Bernardis} P. de Bernardis et al., Nature (London) {\bf 2000}, 404, 955.
	\bibitem{Hannay} S. Hanany et al., Astrophys. J. {\bf 2000}, 545, L5.
	\bibitem{Peebles} P. J. E. Peebles, B. Ratra, Rev. Mod. Phys. {\bf 2003}, 75, 559.
	\bibitem{Padmanabhan} T. Padmanabhan, Phys. Rep. {\bf 2003}, 380, 235.  
	\bibitem{Sahni} V. Sahni, A. Starobinsky, Int. J. Mod. Phys. D {\bf 2000}, 09, 373.
	\bibitem{Clifotn} T. Clifton et al., Phys. Rep. {\bf 2012}, 513, 1.
	\bibitem{Amanullah} R. Amanullah et al., Astrophys. J. {\bf 2010}, 716, 712.
	\bibitem{Komatsu} E. Komatsu et al., Astrophys. J. Suppl. {\bf 2011}, 192, 18. 
	\bibitem{Tegmark} K. Tegmark et al., Phys. Rev. D {\bf 2004}, 69, 103501.
	\bibitem{Nojiri} S. Nojiri, S. D. Odintsov, Int. J. Geom. Methods Mod. Phys. {\bf 2007}, 04, 115.
	\bibitem{Capozziello} S. Capozziello, Int. J. Mod. Phys. D {\bf 2002}, 11, 483.
	\bibitem{Elizalde} E. Elizalde et al., Phys. Rev. D {\bf 2011}, 83, 086006. 
	\bibitem{Bamba} K. Bamba et al., J. Cosmol. Astropart. Phys. {\bf 2011}, 01, 021.
	\bibitem{Houndjo} M. J. S. Houndjo et al., Int. J. Mod. Phys. D {\bf 2017}, 26, 1750024.
	\bibitem{Yousaf1} Z. Yousaf et al., Int. J. Geom. Methods Mod. Phys. {\bf 2018}, 15, 1850146.
	\bibitem{Ilyas} M. Ilyas, Z. Yousaf, M. Z. Bhatti, Mod. Phys. Lett. A {\bf 2019}, 34, 1950082.
	\bibitem{Harko} T. Harko et al., Phys. Rev. D {\bf 2011}, 84, 024020.
	\bibitem{Weyl} H. Weyl, Ann. der Phys. {\bf 1919}, 364, 101.
	\bibitem{Einstein} A. Einstein, Sitzber. Preuss. Akad. Wiss. {\bf 1928}, 17, 217.
	\bibitem{Hayashi} K. Hayashi, T. Shirafuji, Phys. Rev. D {\bf 1979}, 19, 3524.
	\bibitem{Sauer} T. Sauer, Historia Mathematica {\bf 2006}, 33, 399.
	\bibitem{Nester} J.M. Nester, H.-J. Yo, Chin. J. Phys. {\bf 1999}, 37, 113.
	\bibitem{Jimenez} J.B. Jim\'enez, L. Heisenberg, T. Koivisto, Phys. Rev. D {\bf 2018}, 98, 044048.
	\bibitem{Jimenez1} J.B. Jim\'enez, L. Heisenberg, T. Koivisto, J. Cosmol. Astropart. Phys. {\bf 2018}, 2018, 039.
	\bibitem{Heisenberg} L. Heisenberg, Phys. Rep. {\bf 2019}, 796, 1.
	\bibitem{Hohmann} M. Hohmann et al., Phys. Rev. D {\bf 2019}, 99,  024009.
	\bibitem{Soudi} I. Soudi et al., Phys. Rev. D {\bf 2019}, 100, 044008.
	\bibitem{Lazkoz} R. Lazkoz et al., Phys. Rev. D {\bf 2019}, 100, 104027.
	\bibitem{Ambrosio} F. D' Ambrosio et al., Phys. Rev. D {\bf 2022}, 105, 024042.
	\bibitem{Heisenberg1} F. D'Ambrosio et al., Phys. Rev. D {\bf 2022}, 105, 024042.
	\bibitem{Calza} M. Calz\'a, L. Sebastiani, Eur. Phys. C {\bf 2023}, 83, 247. 
	\bibitem{Bajardi} F. Bajardi, D. Vernieri, S. Capozziello, Eur. Phys. J. Plus {\bf 2020}, 135, 912.
	\bibitem{Khyllep} W. Khyllep, A. Paliathanasis, J. Dutta, Phys. Rev. D {\bf 2021}, 103, 103521. 
	\bibitem{Ayuso} I. Ayuso, R. Lazkoz, V. Salzano, Phys. Rev. D {\bf 2021}, 103, 063505.
	\bibitem{Heisenberg2} L. Heisenberg, arXiv:2309.15958v1.
	\bibitem{Barros} B.J. Barros et al., Phys. Dark Universe {\bf 2020}, 30, 100616.
	\bibitem{Jimenez2} J.B. Jim\'enez et al., Phys. Rev. D {\bf 2020}, 101, 103507.
	\bibitem{Anagnostopoulos} F.K. Anagnostopoulos, S. Basilakos, E.N. Saridakis, Phys. Lett. B {\bf 2021}, 822, 136634.
	\bibitem{Flathmann} K. Flathmann, M. Hohmann, Phys. Rev. D {\bf 2021}, 103, 044030.
	\bibitem{Heisenberg3} F. D'Ambrosio, L. Heisenberg, S. Zentarra, arXiv:2308.02250v1.
	\bibitem{Das} A. Das et al., Phys. Rev. D {\bf 2017}, 95, 124011.
	\bibitem{Ghosh1} S. Ghosh et al., Int. J. Mod. Phys. A {\bf 2022}, 35, 2050017.
	\bibitem{Das1} A. Das et al., Nuc. Phys. B {\bf 2020}, 954, 114986.
	\bibitem{Sengupta} R. Sengupta et al., Phys. Rev. D {\bf 2020}, 102, 024037.	\bibitem{Banerjee} S. Banerjee et al., Eur. Phys. J. Plus {\bf 2020}, 135, 185.
	\bibitem{BCP} S. Ghosh et al., J. Cosmol. Astropart. Phys {2021}, 07, 004.
	\bibitem{Yusaf} Z. Yousaf, Phys. Dark Universe {\bf 2020}, 28, 100509.
	\bibitem{Bhatti} M. Z. Bhatti, Z. Yousaf, A. Rehman, Galaxies {\bf 2022}, 10, 40. 
	\bibitem{Pradhan} S. Pradhan et al., Chin. Phys. C {\bf 2023}, 47, 055103.
	\bibitem{Bhatti1} M.Z. Bhatti et al., Indian J Phys {\bf 2023}.
	\bibitem{Bhattacharjee3} D. Bhattacharjee, P. K. Chattopadhyay, J. High Energy Astrophys. {\bf 2024}, 43, 248.
	\bibitem{Sahoo} D. Mohanty, S. Ghosh and P. K. Sahoo, Ann. Phys. (N. Y.) {\bf 2024}, 463, 169636.
	\bibitem{Shamir} M. F. Shamir, Mushtaq Ahmad, Phys. Rev. D {\bf 2018}, 97, 104031.  
	\bibitem{Rahaman2} F. Rahaman et al., Phys. Lett. B {\bf 2012}, 717, 1.
	\bibitem{Chan} R. Chan, M. F. A. da Silva, J. Cosmol. Astropart. Phys. {\bf 2010}, 07, 29.
	\bibitem{Rocha} P. Rocha et al., J. Cosmol. Astropart. Phys. {\bf 2008}, 11, 010.
	\bibitem{Lobo1} F. Lobo et al., Class. Quantum Gravity {\bf 2007}, 24, 1069. 
	\bibitem{Horvat1} D. Horvat, S. Ilijic, A. Marunovic, Class. Quantum Gravity {\bf 2009}, 26, 025003.
	\bibitem{Turimov} B. V. Turimov, B. J. Ahmedov, A. A. Abdujabbarov, Mod. Phys. Lett. A {\bf 2009}, 24, 733.
	\bibitem{Holdom}  B. Holdom, J. Ren, C. Zhang, Phys. Rev. Lett. {\bf 2018}, 120,  222001.
	\bibitem{Zhang1} C. Zhang, Phys. Rev. D {\bf 2020}, 101, 043003.
	\bibitem{Wang} Q. Wang, C. Shi, H. S. Zong, Phys. Rev. D {\bf 2019}, 100, 123003.
	\bibitem{Zhao} T. Zhao et al., Phys. Rev. D {\bf 2019}, 100, 043018.
	\bibitem{Xia} C. J. Xia et al., Phys. Rev. D {\bf 2020}, 101, 103031. 
	\bibitem{Acharya} B. Acharya et al. (MoEDAL Collaboration), Phys. Rev. Lett. {\bf 2021}, 126, 071801.
	\bibitem{Piotrowski} L. W. Piotrowski et al., Phys. Rev. Lett. {\bf 2020}, 125, 091101.
	\bibitem{Zhou} E. P. Zhou, X. Zhou, A. Li, Phys. Rev. D {\bf 2018}, 97, 083015.
	\bibitem{Burgio} G. F. Burgio et al., Astrophys. J., {\bf 2018}, 860, 139. 
	\bibitem{Roupas} Z. Roupas, G. Panotopoulos, I. Lopes, Phys. Rev. D {\bf 2021}, 103, 083015. 
	\bibitem{Horvath} J. E. Horvath and P. H. R. S. Moraes, Int. J. Mod. Phys. D {\bf 2021}, 30, 2150016.
	\bibitem{Harko1} T. Harko, K. S. Cheng, Z. Kovacs, Mon. Not. Roy. Astron. Soc. {\bf 1632}, 400, 1632.
	\bibitem{Ren} J. Ren, C. Zhang, Phys. Rev. D {\bf 2020}, 102, 083003.
	\bibitem{Zhang} C. Zhang, R. B. Mann, Phys. Rev. D {\bf 2011}, 103, 063018.
	\bibitem{Cao} Z. Cao et al., Phys. Rev. D {\bf 2022}, 106, 083007.
	\bibitem{Farhi} E. Farhi, R. L. Jaffe, Phys. Rev. D {\bf 1984}, 30, 2379.
	\bibitem{Fraga} E. S. Fraga, R.D. Pisarski, J. Schaffner-Bielich, Phys. Rev. D {\bf 2001}, 63, 121702.
	\bibitem{Fraga1} E. S. Fraga, A. Kurkela, A. Vuorinen, Astrophys. J. Lett. {\bf 2014}, 781, L25 .
	\bibitem{Alford} M. G. Alford, K. Rajagopal, F. Wilczek, Nucl. Phys.B {\bf 1999}, 537, 443 .
	\bibitem{Rajagopal} K. Rajagopal, F. Wilczek, Phys. Rev. Lett. {\bf 2001}, 86, 3492 .
	\bibitem{Lugones} G. Lugones, J. E. Horvath, Phys. Rev. D {\bf 2002}, 66, 074017.
	\bibitem{Abbott} R. Abbott et al. (LIGO Scientific and Virgo Collaborations), Astrophys. J. Lett. {\bf 2020}, 896, L44 .
	\bibitem{Pretel} J. M. Z. Pretel et al., Phys. Lett. B {\bf 2024}, 848, 138375. 
	\bibitem{Errehymy1} A. Errehymy et al., J. High Energy Astrophys. {\bf 2024}, 44, 410.  
	\bibitem{Tangphati} T. Tangphati et al., Chin. J. Phys. {\bf 2024}, 91, 392.
	\bibitem{Alford1} M. Alford, K. Rajagopal, J. High Energy Phys. {\bf 2002}, 0206, 031.
	\bibitem{Alford2} M. Alford et al., Astrophys. J. {\bf 2005}, 629, 969.
	\bibitem{Weissenborn} S. Weissenborn et al., Astrophys. J. Lett. {\bf 2011}, 740, L14.
	\bibitem{Boehmer} C. G. Boehmer, T. Harko, F.S.N. Lobo, Phys. Rev. D {\bf 2007}, 76, 084014.
	\bibitem{Boehmer1} C. G. Boehmer, T. Harko, F.S.N. Lobo, Class. Quantum Gravity {\bf 2008}, 25, 075016.
    \bibitem{Wang} W. Wang, H. Chen, T. Katsuragawa, Phys. Rev. D {\bf 2022}, 105, 024060.
    \bibitem{Bohmer} C. G. B\"ohmer, A. Mussa, N. Tamanini, Class. Quantum Gravity {\bf 2011}, 28, 245020.
    \bibitem{Israel} W. Israel, Nuo. Cim. B {\bf 1966}, 44, 1.
    \bibitem{Israel1} W. Israel, erratum-ibid. {\bf 1967}, 48, 463.
    \bibitem{Darmois} G. Darmois, Fasticule XXV, Gauthier-Villars, Paris, France, Chap. V (1927).
    \bibitem{Lanczos} K. Lanczos, Ann. Phys. (Berlin) {\bf 1924}, 379, 518.
    \bibitem{Sen} N. Sen, Ann. Phys. (Leipzig) {\bf 1924}, 378, 365.
    \bibitem{Perry} G. P. Perry and R. B. Mann, Gen. Relativ. Gravit. {\bf 1992}, 24, 305.
    \bibitem{Musgrave} P. Musgrave and K. Lake, Class. Quantum Gravity {\bf 1996}, 13, 1885. 
    \bibitem{Buchdahl} H. A. Buchdahl, Phys. Rev. {\bf 1959}, 116, 1027.
    \bibitem{Bhattacharjee1} D. Bhattacharjee, P. K. Chattopadhyay, Chin. J. Phys. {\bf 2025}, 94, 650.
    \bibitem{Madsen} J. Madsen, Lect. Notes Phys. {\bf 1999}, 516, 162.
    \bibitem{Aziz} A. Aziz et al., Int. J. Mod. Phys. D {\bf 2019}, 28, 1941006. 
    \bibitem{Straumann} N. Straumann, Springer, Berlin {\bf 1984}.
    \bibitem{Bohmer} C. G. B\"ohmer, T. Harko, Class. Quantum Gravity {\bf 2006}, 23, 6479.
    \bibitem{Barraco} D. E. Barraco, V. H. Hamity, Phys. Rev. D {\bf 2002}, 65, 124028.
    \bibitem{Ivanov} B. V. Ivanov, Phys. Rev. D {\bf 2002}, 65, 104011. 
\end{thebibliography}


\end{document}